\documentclass[twocolumn,apj]{emulateapj}
\usepackage{apjfonts}
\newcommand{\rotate}{}
\usepackage{lscape}

\shorttitle{Mid- to Far-Infrared Properties of FR\,II Quasars} 
\shortauthors{Fu and Stockton}

\newcommand{\eg}{e.g.,}
\newcommand{\ie}{i.e.,}
\newcommand{\etal}{{et al.}}

\newcommand{\tph}{\phn\phn}
\newcommand{\hph}{\phn\phn\phn}
\newcommand{\spitzer}{{\it Spitzer}}
\newcommand{\um}{$\mu$m}

\newcommand{\kms}{{km s$^{-1}$}}

\newcommand{\msun}{$M_{\odot}$}

\newcommand{\lsun}{$L_{\odot}$}

\newcommand{\leothree}{$L_{\rm [O\,III]}^{\rm eelr}$}

\newcommand{\othree}{{[O\,{\sc iii}]}}

\newcommand{\otwo}{{[O\,{\sc ii}]}}

\begin{document}

\title{FR\,II Quasars: Infrared Properties, Star Formation Rates, and Extended Ionized Gas}
\author{Hai Fu}
\affil{Department of Astronomy, California Institute of Technology, MS 105-24, Pasadena, CA 91125; fu@astro.caltech.edu}
\author{Alan Stockton}
\affil{Institute for Astronomy, University of Hawaii, Honolulu, HI 96822; stockton@ifa.hawaii.edu}

\begin{abstract} 
We present \spitzer\ IRS spectra and MIPS photometry of 12 radio-loud QSOs with FR\,II morphologies at $z \sim 0.3$. Six of the sources are surrounded by luminous extended emission-line regions (EELRs), while the other six do not have such extended nebulae. The two subsamples are indistinguishable in their mid-infrared spectra and overall infrared spectral energy distributions (SEDs). For both subsamples, the mid-infrared aromatic features are undetected in either individual sources or their stacked spectra, and the SEDs are consistent with pure quasar emission without significant star formation. The upper limits to the star formation rate are sufficiently low that starburst-driven superwinds can be ruled out as a mechanism for producing the EELRs, which are instead likely the result of the ejection of most of the gas from the system by blast waves accompanying the launching of the radio jets. The FR\,II quasars deviate systematically from the correlation between host galaxy star formation rate and black hole accretion rate apparently followed by radio-quiet QSOs, implying little or no bulge growth coeval with the current intensive black hole growth. We also present a new \spitzer\ estimate of the star formation rate for the starburst in the host galaxy of the compact steep-spectrum radio quasar 3C\,48. 
\end{abstract}

\keywords{quasars: general --- infrared: galaxies --- galaxies: starburst}

\section{Introduction}\label{intro}

Giant extended emission-line regions (EELRs), with typical radii of 10--30 kpc, are found around roughly half of the low-redshift steep-radio-spectrum quasars with strong nuclear narrow-line emission \citep{Sto87,Fu09}. Although the most luminous EELRs at low redshifts are exclusively associated with steep-spectrum radio-loud quasars, the distribution of the ionized gas typically shows no obvious connection with the radio jets or lobes, or with host galaxy morphologies. These EELRs most likely comprise gas that has been driven out of the host galaxies by superwinds \citep{Sto02,Fu06,Fu07b,Fu08,Fu09}. But are these superwinds mostly produced by starbursts or by the quasars themselves? We have found some evidence in favor of quasar-driven outflows from momentum considerations \citep{Fu06} and from the lack of any optical spectroscopic evidence for a very recent starburst in the radio galaxy 3C\,79, which shows both scattered light from a hidden quasar and a strong EELR \citep{Fu08}. However, to deal with this question more systematically, we need to work with a global star-formation indicator that we can apply to a larger sample of objects.  In particular, for the present program, we have chosen carefully matched samples of steep-radio-spectrum quasars with and without luminous EELRs. Our main goal was to measure or to place strong upper limits to star formation rates (SFRs) in both subsamples of quasars.

A number of measures of SFR are available for galaxies in general. These include optical emission lines, such as H$\alpha$ and [O\,{\sc ii}]\,$\lambda$3727 \citep[\eg][and references therein]{Kew04}, mid-infrared (MIR) fine structure lines, notably [Ne\,{\sc ii}]\,12.8\um\ \citep{Ho07}, polycyclic aromatic hydrocarbon (PAH) emission in the MIR \citep{Sch06,Shi07}, the far-infrared (FIR) continuum \citep[\eg][]{Row97}, and the radio continuum \citep[\eg][]{Con92}. Some of these measures are difficult or impossible to apply to host galaxies of luminous, radio-loud QSOs (\ie\ quasars). For example, photoionization by the quasar UV continuum produces emission lines that will mask those from star formation, and radio emission from the non-thermal radio source can swamp that due to star formation. The PAH emission and the FIR continuum appear to offer the most promise for use with quasar host galaxies, and they have the additional advantage over any optical approaches that they can reveal even deeply obscured starbursts.

In this {\it Paper}, we present and discuss \spitzer\ MIR spectra and FIR photometry of 13 steep-radio-spectrum and/or FR\,II \citep{Fan74} quasars at $z \sim 0.3$. Throughout we assume a cosmological model with $H_0=70$ km s$^{-1}$ Mpc$^{-1}$, $\Omega_m=0.3$, and $\Omega_{\Lambda}= 0.7$.

\begin{deluxetable*}{lcccccccccccc}
\rotate
\setlength{\tabcolsep}{.11cm}
\tablewidth{0pt}
\tabletypesize{\footnotesize}
\tablecaption{Quasar Sample \label{tab1}}
\tablehead{ 
\colhead{} & \colhead{} &\colhead{} &
\colhead{log \leothree} & \colhead{log $\lambda L_{4861}$} & \colhead{} & 
\colhead{log $P_{\rm 178}^{\rm tot}$} & \colhead{Size} & 
\colhead{$F_{\rm 24 \mu m}$} & \colhead{$F_{\rm 70 \mu m}$} & \colhead{$F_{\rm 160 \mu m}$} & 
\colhead{log $\lambda L_{\rm 60 \mu m}$} & \colhead{log $L_{\rm PAH}$}
\\
\colhead{Name} & \colhead{Designation} & \colhead{$z$} & 
\colhead{(erg s$^{-1}$)} & \colhead{(erg s$^{-1}$)} & \colhead{$\alpha_{\nu}$} & 
\colhead{(W Hz$^{-1}$)} & \colhead{(kpc)} & 
\colhead{(mJy)} & \colhead{(mJy)} & \colhead{(mJy)} &
\colhead{(erg s$^{-1}$)} & \colhead{(erg s$^{-1}$)}
\\
\colhead{(1)} & \colhead{(2)} & \colhead{(3)} & 
\colhead{(4)} & \colhead{(5)} & \colhead{(6)} & 
\colhead{(7)} & \colhead{(8)} & \colhead{(9)} & 
\colhead{(10)} & \colhead{(11)} & \colhead{(12)} &
\colhead{(13)}  
}
\startdata
\multicolumn{13}{c}{CSS Quasar with EELR}   \\
\hline
        3C\,48&0137$+$3309&0.367&~~\,42.75&45.11&0.82&28.41&  \hph3&   128.6&   754.6&  ~475.9&46.11&~~\,43.97\\
\hline
\multicolumn{13}{c}{FR\,II Quasars with EELRs}   \\
\hline
     3C\,249.1&1104$+$7658&0.312&~~\,42.82&45.52&0.84&27.63&\phn137&\phn43.9&\phn60.2& $<$14.0&44.84& $<$42.70\\
      Ton\,616&1225$+$2458&0.268&~~\,42.59&44.63&0.80&26.67&\phn279&\phn11.9&\phn12.6& $<$13.5&44.04& $<$42.40\\
      Ton\,202&1427$+$2632&0.366&~~\,42.32&45.19&0.62&26.94&   1169&\phn40.3&\phn61.0& $<$11.5&44.99& $<$42.81\\
     4C\,37.43&1514$+$3650&0.371&~~\,42.97&45.27&0.80&27.36&\phn276&\phn30.0&\phn35.0& $<$13.0&44.76& $<$42.84\\
     3C\,323.1&1547$+$2052&0.264&~~\,42.20&45.18&0.71&27.33&\phn285&\phn32.4&\phn18.8& $<$14.1&44.19& $<$42.62\\
  PKS\,2251+11&2254$+$1136&0.326&~~\,42.47&45.17&0.73&27.37& \tph47&\phn42.2&\phn30.1& $<$26.6&44.59& $<$42.72\\
\hline
\multicolumn{13}{c}{FR\,II Quasars without EELRs}   \\
\hline
     4C\,25.01&0019$+$2602&0.284& $<$40.95&45.11&0.66&26.82&\phn204&\phn67.8&   116.4&~~\,53.6&45.06& $<$42.82\\
     4C\,13.41&1007$+$1248&0.241& $<$40.38&45.38&0.67&26.95&\phn387&\phn72.9&\phn86.5& $<$22.1&44.78& $<$42.78\\
       3C\,246&1051$-$0918&0.344& $<$40.92&45.21&0.78&27.68&\phn415&\phn20.5&\phn20.0& $<$15.8&44.46& $<$42.56\\
     4C\,49.22&1153$+$4931&0.334& $<$40.40&44.71&0.49&27.32& \tph80&\phn25.3&\phn42.2& $<$13.6&44.75& $<$42.71\\
       3C\,351&1704$+$6044&0.372& $<$41.29&45.40&0.74&27.85&\phn282&\phn99.2&   173.7&~~\,68.2&45.47& $<$43.08\\
     4C\,31.63&2203$+$3145&0.295& $<$40.94&45.41&0.16&26.99&\phn352&\phn67.2&\phn83.3&~~\,54.2&44.95& $<$42.81     
\enddata
\tablecomments{
Col.\,(1): Common name. 
Col.\,(2): J2000.0 designation.
Col.\,(3): Redshift.
Col.\,(4): Extended \othree\ $\lambda5007$ luminosity in logarithmic and
upper limits for non-EELR quasars.
Col.\,(5): Optical continuum luminosity under H$\beta$.
Col.\,(6): Radio spectral index ($f_{\nu} \propto \nu^{-\alpha_{\nu}}$)
between 0.4 and 2.7 GHz.
Col.\,(7): 178 MHz radio luminosity in logarithmic (The value for
Ton\,202 was interpolated between 151 and 365 MHz fluxes).
Col.\,(8): Projected linear size of the radio source (References: \citealt{Aku91,Bri94,Fen05,Gop00,Gow84,Kel94,Sto85}).
Cols.\,(9)--(11): Observed flux densities in the MIPS 24, 70, and 160 \um\ bands.
Col.\,(12): Interpolated luminosity at rest wavelength 60 \um\ from MIPS photometry.
Col.\,(13): Observed luminosity and upper limits of the PAH 7.7$\mu$m feature from IRS spectra.
The reference for Cols.\,(4)--(6) is \citet{Sto87}. Data in Col.\,(7) are compiled from the NASA/IPAC Extragalactic Database (NED).  
}
\end{deluxetable*}

\section{Sample Selection and \spitzer\ Observations}

Our sample of quasars has been selected from the \citet{Sto87} survey for extended emission around low-redshift QSOs. We have a principal sample of 12 FR\,II quasars (which we will refer to as ``the FR\,II sample''), half of which have luminous EELRs and half of which do not, to fairly strong upper limits. The EELR and non-EELR quasars have been paired as closely as possible in both redshift and rest-frame optical continuum luminosity. Characteristics of all of the objects are tabulated in Table~\ref{tab1}.  For the EELR and non-EELR quasars, the mean redshifts are 0.318 and 0.312, respectively. The continuum luminosities are measured at H$\beta$, and the mean values for log $\lambda L_{4861}$ (erg s$^{-1}$) are 45.23 and 45.26. The radio powers of the subsamples are also similar, with mean values for log $P_{178 {\rm MHz}}$ (W Hz$^{-1}$) of 27.32 and 27.44. We also include 1 EELR quasar with a compact steep-spectrum (CSS) one-sided radio jet, 3C\,48, which is a special object in a number of other ways (see, e.g., \citealt{Can00a,Sto07}) and will be discussed apart from the FR\,II sample.

Deep MIR spectra of all 13 quasars were obtained with the \spitzer\ Infrared Spectrograph \citep[IRS;][]{Hou04}  as part of our Cycle 4 GO program (PID 40001). All of the objects were observed in the standard staring mode with the Short Low 1st order module (SL1, 7.5$-$14.2 \um). The slit width was 3.7\arcsec\ or 16.5 kpc at $z = 0.3$. Each observation consisted of a 60 s ramp for 8$-$16 cycles with two nod positions. The total integration time was about 16$-$32 minutes. We used the IDL procedure IRSCLEAN\footnote{All of the data reduction software mentioned in this section is available at the \spitzer\ Science Center website at http://ssc.spitzer.caltech.edu/.} (ver. 1.9) to remove bad pixels in the pipeline post-BCD coadded nod-subtracted images and extracted the spectra with SPICE (ver. 2.1.2) using the optimal extraction method for point sources. 

FIR photometry at 24, 70, and 160 \um\ was obtained with the Multiband Imaging Photometer for \spitzer\ \citep[MIPS;][]{Rie04} either as part of our GO program or from the archive (PIDs 00049, 00082, 20084). All of the observations were obtained using the small-field default resolution photometry mode. Typical integration time per pixel was 42 s at 24 \um\ (3 s exposure for 1 cycle or 3 s$\times$1), 700 s at 70 \um\ (10 s$\times$7), and 80 s at 160 \um\ (10 s$\times$4). Our data reduction started from the pipeline BCD files. For 24 \um\ observations, final mosaic images were constructed with the MOPEX software after self-flat-fielding and background correction. At 70 and 160 \um, we used the pipeline filtered BCDs to construct mosaics for most of the sources. 3C\,48 and 3C\,351 are so bright at 70 \um\ that they have to be masked out before applying the spatial and temporal filters; we thus used the offline IDL filtering procedures by D.~Fadda and D.~Frayer to generate filtered BCDs. The non-filtered 160 \um\ BCDs of 3C\,48 were used because they result in a better mosaic image than the filtered BCDs. Point sources were extracted with APEX, and point response function (PRF)-fitting photometry was carried out. All of the sources were detected at 24 and 70 \um, but only 4 were detected at 160 \um. Upper limits at 160 \um\ were estimated from the standard deviation images associated with the mosaics using an aperture of 48\arcsec\ and corrected for the finite aperture size by multiplying the values by 1.6 (MIPS Data Handbook).  The photometry results are listed in Table~\ref{tab1}. The systematic uncertainties were estimated to be 4\% at 24 \um, 5\% at 70 \um, and 12\% at 160 \um\ \citep{Eng07,Gor07,Sta07}. Statistical errors from the PRF fitting are much smaller and therefore could be neglected.

\begin{figure*}[!tb]
\epsscale{1.1}
\plotone{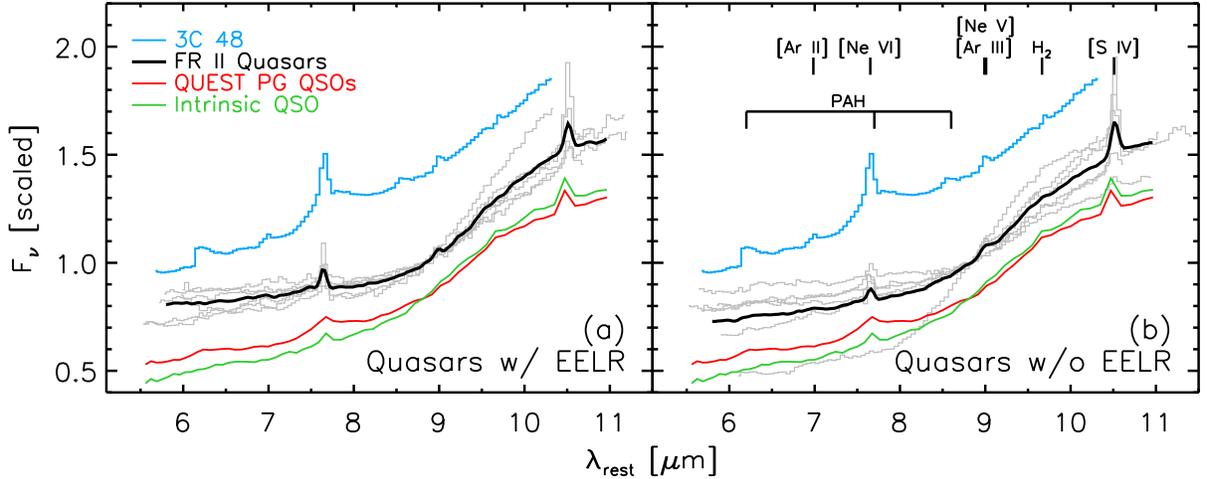}
\caption{
Average IRS spectrum (thick black line) of ({\it a}) six EELR quasars and ({\it b}) six non-EELR quasars. Individual spectra have been normalized at 8.8 \um\ and are shown as light grey curves. Overplotted for comparison are the IRS spectrum of 3C\,48 (blue), the average spectrum of 28 PG QSOs (red), and the average ``intrinsic" QSO spectrum derived from 8 FIR-weak PG QSOs after subtraction of a starburst component (green; \citealt{Net07}); all have been arbitrarily offset for clarity. Important emission features are labelled in panel {\it b}.
\label{fig:irs}} 
\end{figure*}

\begin{figure*}[!tb]
\epsscale{0.5}
\plotone{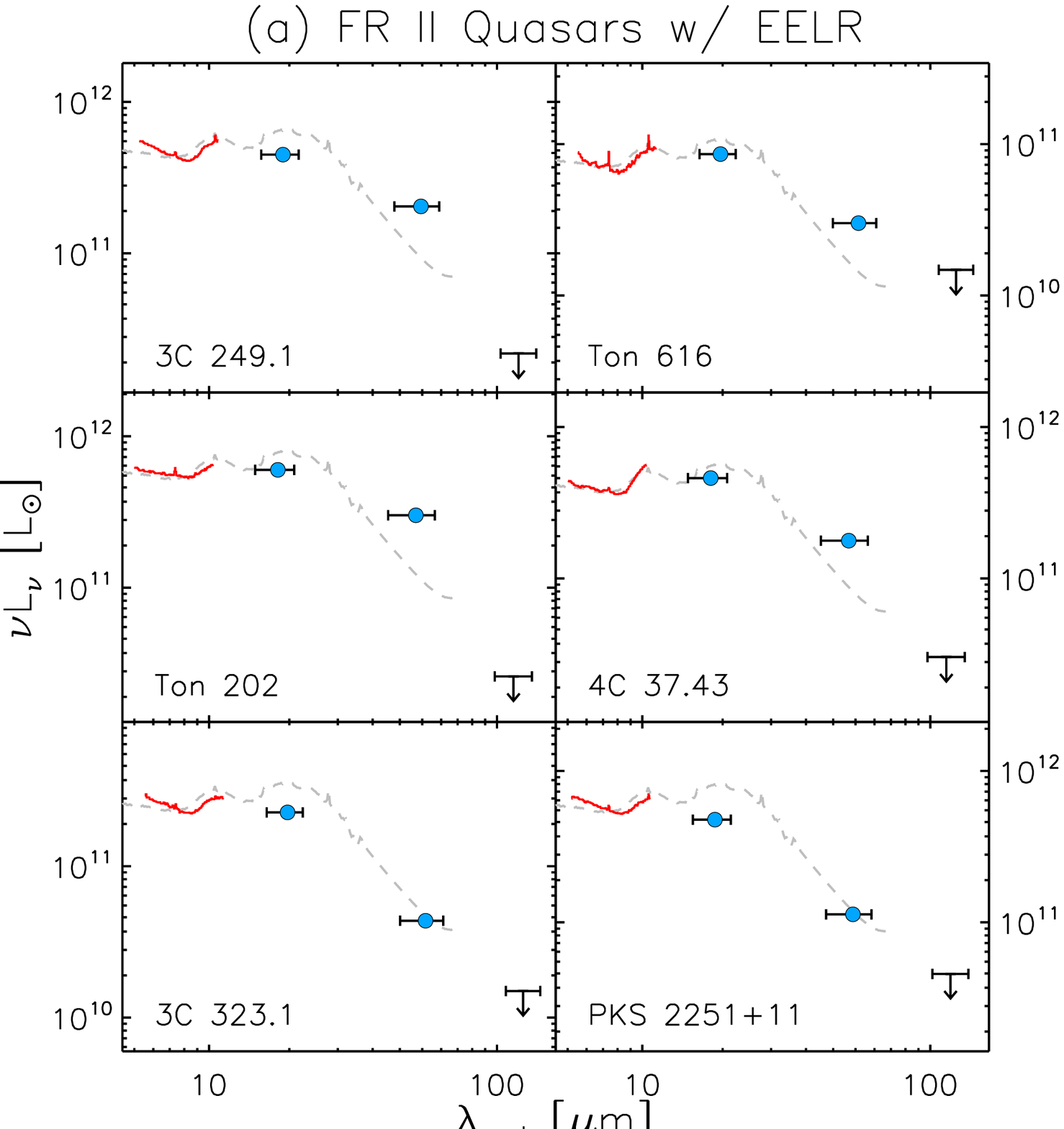}
\hskip 0.2in
\plotone{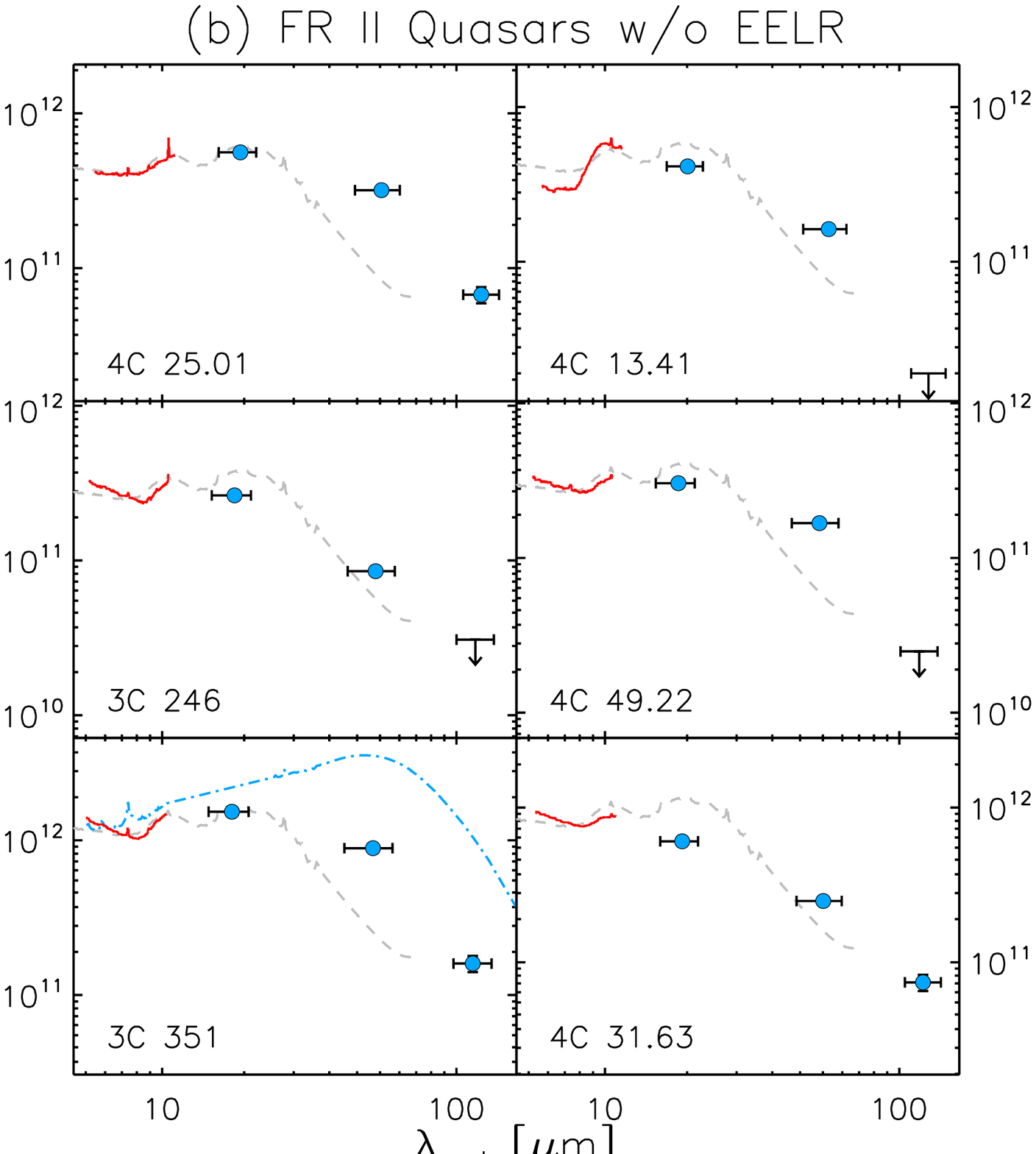}
\caption{
MIR to FIR SEDs of FR\,II quasars: ({\it a}) EELR quasars and ({\it b}) non-EELR quasars. The red curves are the IRS spectra and the data points (detections) and downward arrows (upper limits) are MIPS photometry. Overplotted for comparison is the average ``intrinsic" QSO SED, scaled to the IRS luminosity of each source \citep[grey dashed curve;][]{Net07}. In the panel for 3C\,351, the most FIR luminous FR\,II quasar in our sample, we also show the SED of 3C\,48 as the blue dash-dot line (no scaling has been performed).  
\label{fig:sed}} 
\end{figure*}

\begin{figure*}[!tb]
\epsscale{1.1}
\plottwo{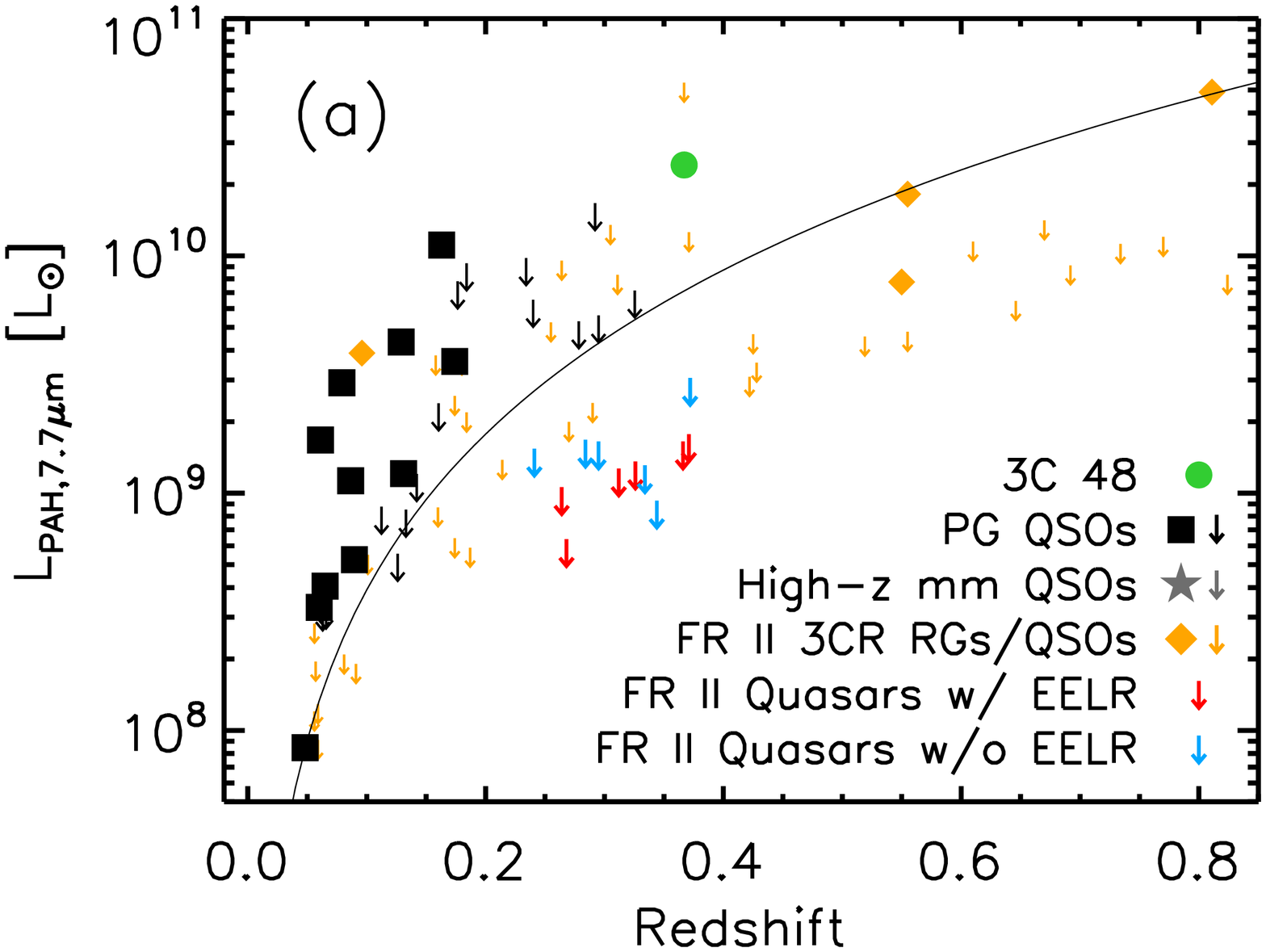}{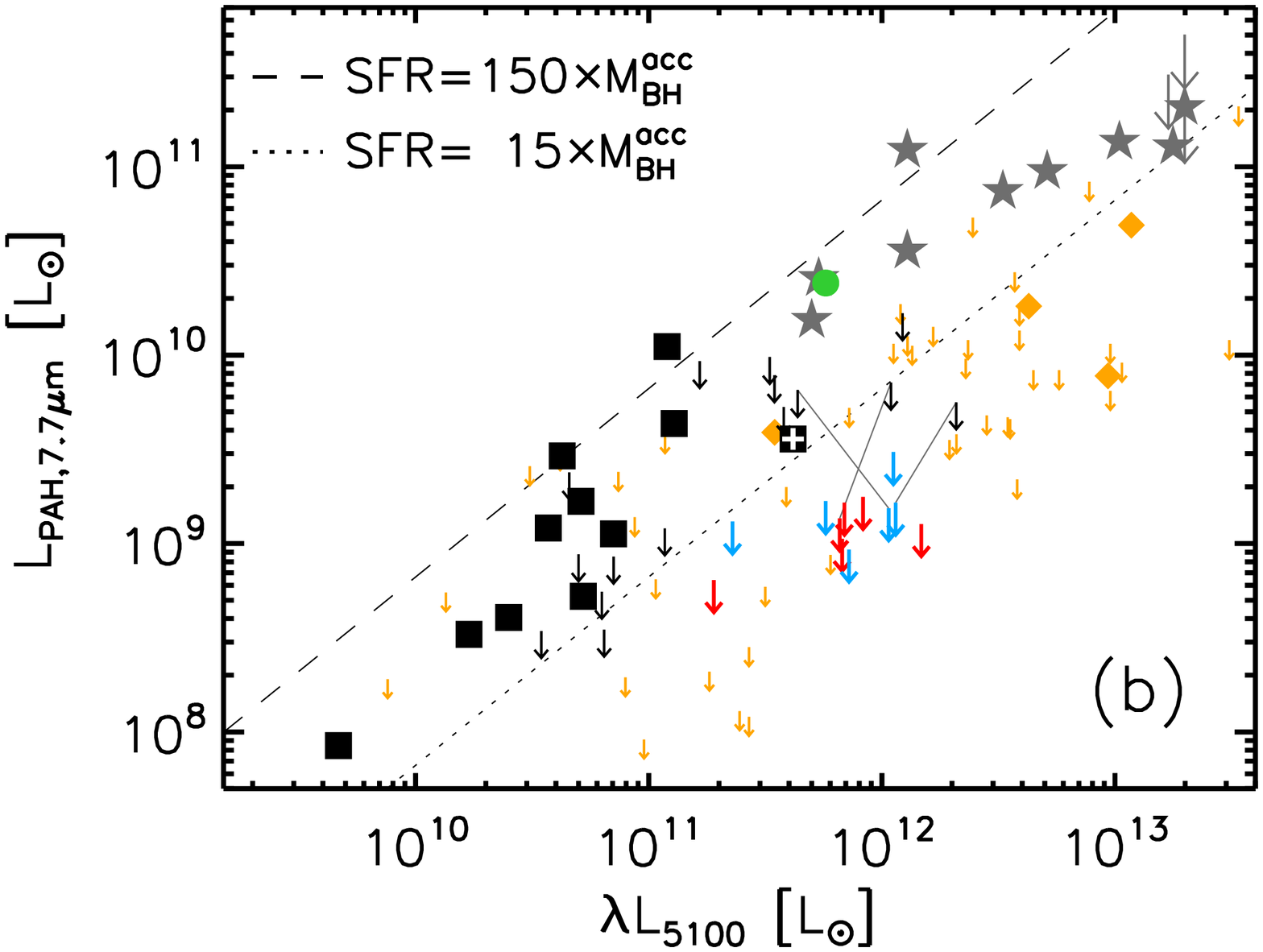}
\caption{
({\it a}) PAH 7.7 \um\ luminosity vs. redshift for various
samples, as indicated.  The solid curve indicates a constant flux limit of
$6\times10^{-14}$ erg s$^{-1}$ cm$^{-2}$, approximating that of the PG
sample. Note that the upper limits for our FR\,II
quasar sample (larger red and blue arrows) are below or comparable to
the detections for the PG QSO sample, in spite of the higher redshift
of the FR\,II sample. The symbols given in the legend for this panel
apply to panel $b$ as well.
({\it b}) PAH 7.7 \um\ luminosity vs. optical continuum luminosity.
The PG QSOs from \citet{Sch06} are shown as filled black squares
(detections) or downward black arrows (upper limits). Similarly, the
high-redshift mm-bright QSOs from \citet{Lut08} are shown as gray
stars or large gray downward arrows, and FR\,II 3CR quasars and radio
galaxies from \citet{Shi07} are shown as orange diamonds or downward
arrows. Our FR\,II quasars with EELRs are shown as downward red
arrows, and the ones without EELRs are downward blue arrows. 3C\,48 is
indicated by a green filled circle. The only FR\,II quasar that is in the PG sample
but not in our sample, PG\,2349$-$014, is marked by a filled black
square with a white plus sign. The dotted and dashed lines show
constant ratios of star formation rates and black hole accretion
rates.  The three common objects between our sample and the PG QSOs
are connected with grey solid lines. The $\sim4\times$ reduced upper
limits of $L_{\rm PAH}$ for the three common objects result from our
much longer IRS integrations.
The FR\,II quasars are clearly under-luminious in the PAH emission for
their optical luminosity (which can be considered to be a proxy for
black-hole accretion rate) with respect to both the PG QSOs that are
not FR\,II quasars and the mm-bright QSOs.
\label{fig:opt}}
\end{figure*}

\section{Results}\label{results}

We did not detect PAH emission in any of the FR\,II quasars, although strong PAH features were clearly seen at 6.2, 7.7, and 8.6 \um\ in the spectrum of the CSS quasar 3C\,48. It was already known that the host galaxy of 3C\,48 is undergoing intense star formation, with evidence from both its warm {\it IRAS} FIR color and the young stellar populations revealed by deep optical spectra \citep{Can00a}. The luminosity of the PAH 7.7 \um\ feature ($L_{\rm PAH}$) is about $2.4\times10^{10}$ \lsun, implying a SFR of 680 \msun\ yr$^{-1}$ (SFR[\msun\ yr$^{-1}$] = 283 $L_{\rm PAH}$[10$^{10}$ \lsun], assuming $L_{\rm PAH}/L_{\rm IR}$ = 0.0061 and SFR[\msun\ yr$^{-1}$] = 1.73 $L_{\rm IR}$[10$^{10}$ \lsun]; \citealt{Vei09,Ken98}. Here, and throughout this paper, $L_{\rm IR}$ refers to the {\it star-forming} luminosity in the rest-frame 8--1000 $\mu$m range). The aromatic features remain undetected even in the stacked spectrum of the FR\,II sample. In Fig.~\ref{fig:irs} we show average spectra of two subsamples---FR\,II quasars with and without EELRs, each containing six objects. The average spectra of the two subsamples do not differ: basically, one sees only narrow high-ionization emission lines superposed on a featureless continuum. \citet{Net07} provided a composite spectrum of 28 PG QSOs with IRS spectra, as well as an ``intrinsic" QSO spectral energy distribution (SED) from 1.2 to 70 \um, which was constructed by combining 8 FIR-weak PG QSOs (presumably with little star formation) and subtracting a scaled starburst component. The MIR portion of these two SEDs is also shown in Fig.~\ref{fig:irs} for comparison. Broad PAH emission is clearly seen in the composite PG QSO spectrum, but by design it is absent from the intrinsic QSO spectrum. The average spectra of FR\,II quasars are consistent with being entirely intrinsic QSO emission. 

We compute the upper limits to the PAH 7.7\um\ flux based on the noise in the region between 6.5 and 8.5 \um\ excluding the narrow [Ne\,{\sc vi}] line. A low-order polynomial fit to the continuum is subtracted from each individual spectrum and the standard deviation (\ie\ noise) of the residual is determined. To obtain the upper limits, we assume that the undetected PAH has a Lorentzian profile with the same width (0.6 \um\ FWHM) as our best fit to the 3C\,48 PAH 7.7\um\ feature and that its peak amplitude cannot be greater than 5 times the noise level. This peak-to-noise ratio of 5 corresponds to a total S/N of $\sim$14 if we use the optimal aperture of 1.45$\times$ FWHM. The PAH upper limits reported in Table~\ref{tab1} are consistent with the upper limits obtained by \citet{Sch06} for the 3 quasars that are in both samples, once one takes into account the $\sim$18$\times$ longer integrations of our observations. Furthermore, we found that the PAH feature would have been clearly seen in both stacked spectra and individual spectra if synthetic PAH emission with these upper limits were to be added to every object. 

Figure~\ref{fig:sed} shows the MIR--FIR SEDs of our FR\,II quasars. A scaled version of the \citet{Net07} intrinsic QSO SED is included in each panel for comparison. Most of the SEDs probed by IRS show steady rises towards both ends of the spectrum and a dip near 8 \um; these features are readily seen in the intrinsic QSO SED and can be attributed to a combination of hot dust blackbody emission and silicate emission that peaks near $\sim$10 \um. As an example of starburst quasars, the SED of 3C\,48 is overplotted in the panel of 3C\,351, the most FIR luminous FR\,II quasar in our sample. The SEDs of the FR\,II quasars can be mostly explained as intrinsic QSO emission without a starburst component. The SEDs of quasars with EELRs and the ones without EELRs are essentially the same. These results are consistent with our PAH measurements, as described in the two preceding paragraphs.

A luminosity correlation between the optical continuum and the PAH emission has been observed in low-redshift PG QSOs and high-redshift millimeter-bright QSOs \citep{Net07,Lut08}. This correlation implies an intimate connection between QSO nuclear activity, as indicated by the strength of the continuum at rest-frame 5100 \AA, and star formation in their host galaxies, as traced by the PAH emission. While the median redshift of our sample is significantly above that of the PG sample, our IRS exposures, which are typically $\sim18$ times those for the PG sample, are sufficient to give luminosity upper limits that are lower than the detections or upper limits for the bulk of the PG sample (see Fig.~\ref{fig:opt}$a$). Figure~\ref{fig:opt}$b$ examines whether our FR\,II quasars are consistent with this optical-continuum/PAH correlation. We converted the continuum luminosities under H$\beta$ from \citet{Sto87} to $\lambda L_{5100}$ by multiplying the values by a factor of 1.74 (after the correction for cosmology), which was determined by comparing the two continuum luminosities of the three quasars in both our sample and the \citet{Net07} sample. 
Also plotted in Figure~\ref{fig:opt} are the 32 FR\,II radio galaxies and 15 FR\,II quasars from the 3CR sample of \citet{Shi07}, which have both 7.7 \um\ PAH measurements from IRS spectra and \othree\,$\lambda5007$ and/or \otwo\,$\lambda3727$ fluxes from the literature (see \citealt{Jac97} and references therein\footnote{An electronic version of the catalog is available at http://www.science.uottawa.ca/$\sim$cwillott/3crr/3crr.html}). We converted their \othree\ or \otwo\ luminosities into $ \lambda L_{5100}$ based on the observed \othree-continuum correlation for type-1 QSOs \citep[$ \lambda L_{5100} \simeq 320\times L_{\rm [O III]}$;][]{Zak03,Hec04} and assuming \othree/\otwo\ = 3.3, as measured from sources where fluxes of both lines are available. 
It is evident that the FR\,II quasars/radio galaxies fall systematically below the correlation established by the PG QSOs and the \citet{Lut08} mm-bright QSOs.

Five of the QSOs in the \citet{Net07} sample are radio-loud, among which three are also in our sample (4C\,13.41, 4C\,31.63, and PKS\,2251+11). The remaining two are PG\,1302$-$102 and PG\,2349$-$014. PG\,1302$-$102 has a flat radio spectrum and PAH was not detected ($L_{\rm PAH} <  5.3\times10^9$ \lsun; \citealt{Sch06}). PG\,2349$-$014 has a steep radio spectrum and an FR\,II radio morphology. Although the aromatic 7.7\um\ feature has been claimed to be detected in PG\,2349$-$014 ($L_{\rm PAH} = 3.6\times10^9$ \lsun; \citealt{Sch06}), the low PAH luminosity places it at the transitional region between the FR\,II sources and the radio-quiet QSOs in Fig.~\ref{fig:opt}{\it b} (the black square with a white plus sign). 

The optical continuum luminosity at 5100 \AA\ is widely used to estimate the bolometric luminosity for type-1 AGN ($L_{bol} \sim 10 \lambda L_{5100}$), which in turn provides the black hole accretion rate (BHAR) given a typical radiative efficiency of $\eta = 0.054$ \citep{Mart08}. The PAH luminosity can be converted into a SFR. As shown in Fig.~\ref{fig:opt}$b$, most of the PG QSOs in the \citet{Net07} sample and the mm-bright QSOs in the \citet{Lut08} sample are within a narrow range of SFR/BHAR ratios, 15 $<$ SFR/BHAR $<$ 150; while our FR\,II quasars show significantly lower SFR at any given BHAR (we have conservatively assumed that $L_{\rm PAH} = 0.0061 L_{\rm IR}$; see discussion of this ratio in \S \ref{sf_eelr}). 
As the referee has pointed out, the observed separation in Figure~\ref{fig:opt}{\it b} could also be explained by an on-average 10$\times$ higher radiative efficiency, $\eta$, for FR\,II sources than for radio-quiet QSOs. However, the theoretically allowed range of $\eta$ is only from 0.054 to 0.42 (non-rotating and maximally rotating black holes, respectively; \citealt{Sha83}), and, to make it worse, the presence of a jet can {\it reduce} the efficiency \citep{Jol08}. Therefore, a difference in $\eta$ could, at best, partially explain the large offset that we observe in Figure~\ref{fig:opt}{\it b}. 
We note that to explain the Magorrian bulge mass$-$black hole mass relation, under the assumption that star formation and black hole accretion are always coeval, a SFR/BHAR $\simeq$ 700 is required \citep{Har04}. The measured low values have been used to argue for a much longer duration for star formation than that for the AGN activity \citep{Net07}, making the existence of a SFR-BHAR correlation questionable.

\section{Implications}

\subsection{Can Quasar EELRs be Produced by Starburst Superwinds?}\label{sf_eelr}

Most of our EELR quasars have upper limits to  $L_{\rm PAH}$ of less than $2\times10^9$ \lsun. The non-detection of PAH in the stacked spectrum of the entire sample of 12 FR\,II quasars shows that such upper limits are quite conservative. These upper limits imply a SFR $<12$ \msun\ yr$^{-1}$. In the above calculation we used a higher $L_{\rm PAH}/L_{\rm IR}$ ratio of 0.028 than in calculating the SFR of 3C\,48 (0.0061, \S~3), because it is both observed \citep[\eg][]{Sch06} and theoretically predicted \citep[\eg][]{Dal02} that $L_{\rm PAH}/L_{\rm IR}$ increases with decreasing IR luminosity. Following \citet{Shi07}, we determined $L_{\rm PAH}/L_{\rm IR} \simeq 0.028$ for $L_{\rm PAH} = 2\times10^9$ \lsun\ from the star-forming templates of \citet{Dal02}. However, as discussed in \citet{Fu06}, to inject enough momentum into a typical EELR within a dynamical timescale of 10 Myr to explain the observed chaotic velocity field, the SFR of the central star burst must exceed $\sim60$ \msun\ yr$^{-1}$ (eqs. [2]$-$[3] in \citealt{Vei05}). More importantly, the absence of significant PAH emission in both FR\,II quasars with EELRs and the ones without strongly suggests that starburst is not a critical factor in the formation of an EELR. This conclusion is also supported by the MIPS photometry, which indicates that the FIR SED is dominated by warm dust directly heated by the quasars (Fig.~\ref{fig:sed}).

\subsection{Star Formation in FR\,II Quasars}

Relations among mergers, starbursts and nuclear activities have long been sought after by extragalactic researchers \citep[\eg][]{San88}. Such connections are attractive because starbursts and AGN require many common triggering conditions, and gas-rich mergers can in principle provide these necessary conditions. Optical spectroscopy of QSO host galaxies, although technically challenging, has revealed diverse starforming properties. While active/recent star formation ($< 300$ Myr) have been detected in the host galaxies of certain classes of QSOs, such as FIR-loud QSOs and post-starburst QSOs \citep[e.g.,][]{Can01}, the majority of optically luminous QSOs show signs of past merger events and massive star formation that happened 1$-$2 Gyr ago (see \citealt{Can06} for a review; also \citealt{Can07,Ben08}). A similar result has been found for the FR\,II radio galaxy 3C\,79, which shows both a luminous EELR and evidence for a luminous quasar hidden from our direct view \citep{Fu08}.

About 40\% of the PG QSOs studied by \citet{Sch06} show PAH emission in individual \spitzer\ IRS spectra, and PAH is also present in the average spectrum of the remaining QSOs. In sharp contrast, our considerably deeper IRS spectra of steep-spectrum radio-loud quasars show no evidence for PAH emission, either individually or averaged, except in the case of the CSS quasar 3C\,48. In fact, previous \spitzer\ observations have found that powerful radio galaxies and quasars from the 3CR catalog \citep{Spi85} generally show FIR colors consistent with hot dust directly heated by the AGN \citep{Shi05} and weak or undetectable PAH emission \citep{Haa05,Ogl06,Cle07,Shi07}, suggesting a lower level of star formation compared to the QSOs from the PG and 2MASS samples \citep{Sch83,Cut01}. The fraction of 3CR quasars/radio galaxies that show detectable PAH features is only 18\%, compared to $\sim$48\% for both PG and 2MASS QSOs in the same redshift range ($z < 0.5$) and observed with the same instrument.

There are substantial reasons for believing that radio-loud quasars, at least at low redshifts, are drawn from a population of massive elliptical galaxies that may have suffered mergers in the fairly recent past, while many radio-quiet QSOs are found either in disk galaxies or in ongoing gas-rich mergers \citep{Sik07,Sik08,Wol08}. We can test this explanation by looking at the PG QSOs in the \citet{Sch06} sample that also have morphological determinations by \citet{Guy06} or elsewhere in the literature. Of the 9 objects which had individual PAH detections and for which morphological information is available (8 of which are classified by Guyon et al.), all are classified either as having a disk or as ``strongly interacting,'' a term that almost always implies a gas-rich interaction or merger. It is not at all surprising that such host galaxies should show substantial star formation. 

\subsection{How Does 3C\,48 Fit In?}

At first sight, the CSS quasar 3C\,48 seems to be a counter-example to the trend both \citet{Shi07} and we have found for steep-radio-spectrum quasars. In particular, 3C\,48 has one of the most luminous EELRs among low-redshift quasars, yet its host galaxy is also undergoing star formation at a prodigious rate (\S \ref{results}). Is 3C\,48 simply an anomaly, perhaps related to its status as a CSS source, or can it be understood within the framework we have described? Because of the small size of our sample and the rather large uncertainties in quasar lifetimes and duty cycles, we can do little more than explore some of the possibilities related to these options. First, however, we summarize some of the special characteristics of 3C\,48 and its host galaxy (see \citealt{Sto07} for more detail).

Because 3C\,48 shows a one-sided jet, one might suspect that the jet is oriented almost along our line-of-sight and is strongly Doppler boosted; however, \citet{Wil91} have argued from the weakness of the nuclear radio component that the jet is oriented closer to the plane of the sky and that the radio structure is intrinsically one sided; so the projected jet length of $\sim3$ kpc is probably not too far from its actual length. 3C\,48 is one of the only two CSS sources among powerful quasars with $z<0.5$. Such sources are rare, not because they are intrinsically uncommon, but because the CSS stage of an extended radio source likely lasts only a short time, typically $\sim10^4$ years \citep{deS99}. The host galaxy of 3C\,48 is clearly undergoing a major merger \citep{Sto87,Can00a,Sch04,Sto07}. The evidence suggests that the current episode of quasar activity has been triggered only relatively recently. 

We consider first the possibility that 3C\,48 is simply an earlier stage of the formation of an EELR like those in our FR\,II EELR subsample. This picture has considerable appeal, because we do see a massive ($\gtrsim10^9$ M$_{\odot}$) outflow of ionized gas extending over a large solid angle from a region near the base of the CSS radio jet, with velocities ranging up to $\sim1000$ \kms \citep{Cha99,Can00a,Sto07}. This outflow is consistent with the sort of quasar-jet-driven wind we have supposed to be responsible for producing the large-scale EELRs seen in 3C\,48 and other quasars. Nevertheless, there are difficulties with this simple evolutionary view. Since 3C\,48 also has an EELR on scales much larger than that of the current radio jet, one would have to appeal to a previous episode of quasar activity in order to produce it according to the scenario we have outlined.  It is also not clear that the respective lifetimes of the EELR and the starburst would allow evidence for the starburst to die out quickly enough. EELR lifetimes are likely on the order of $10^7$ years, both from their close association with radio sources and from dynamical estimates. Vigorous star formation in 3C\,48 is taking place not only in the nuclear region, but also over most of the host galaxy \citep{Can00a}. While we do not have similarly detailed spectroscopy for any of the host galaxies in our FR\,II sample, our deep spectroscopy of the FR\,II EELR radio galaxy 3C\,79 has shown that its last significant episode of star formation was at least 1 Gyr ago \citep{Fu08}. This EELR, at least, could not have been produced in conjunction with a starburst like that occurring in 3C\,48. While it is clear that some QSOs are triggered more-or-less simultaneously with massive starbursts \citep[e.g.,][]{Can01}, many showing similar levels of QSO activity seem to have been triggered $\sim1$ Gyr {\it after} major starbursts \citep{Can06,Can07,Ben08}, although we do not yet have a clear understanding why this should be so.

It seems, then, that 3C\,48 is not likely to be an example of a direct precursor to objects like those in our FR\,II EELR subsample. Nevertheless, it does seem to be a close cousin. Its EELR has a luminosity similar to those of our other EELR quasars, and the massive outflow of ionized gas from the base of the radio jet seems a striking confirmation of the sort of process we have inferred, from less direct evidence, to have produced the massive EELRs in our FR\,II subsample \citep[][and references therein]{Fu09}.  The crucial difference between 3C\,48 and the rest is in whether a massive starburst is triggered approximately contemporaneously with the formation of the EELR. This difference, in turn, is likely to depend on the details of the mechanism by which gas is supplied to the system. In particular, the uniformly and abnormally low metallicities of the broad-line gas in the EELR quasars for which this measurement can be made \citep{Fu07a} indicate an {\it external} source for this gas and suggest the mergers of a gas-rich late-type galaxies with massive early-type galaxies with little cold gas of their own. Although we do not have direct metallicity information for the broad-line region in 3C\,48, the morphology of the merger, the large amounts of molecular gas present \citep{Win97}, and the fact that it has $L_{\rm IR} > 10^{12}$ $L_{\odot}$, qualifying it as an ``ultra-luminous infrared galaxy,'' all suggest that it is a merger between two roughly equal-mass gas-rich galaxies.

\subsection{Overview}

Our deep {\it Spitzer} IRS spectra and MIPS photometry of matched subsamples of FR\,II quasars with and without luminous EELRs give tight upper limits to current SFRs for all of the quasars in both subsamples, supporting and extending earlier results on FR\,II quasars and radio galaxies by \citet{Shi07}. These upper limits indicate that SFRs, relative to BHARs, are generally much lower in FR\,II quasars than they are in optically selected QSO samples. In the FR\,II quasars, then, very little bulge mass is being added via star formation during the current episode of black-hole growth. Our star-formation upper limits are also sufficiently low that we can discount the possibility of galactic-scale superwinds resulting from star formation in the host galaxies of these quasars. As we mentioned in \S \ref{intro}, the most luminous EELRs are always associated with quasars with strong radio jets, yet there is little significant morphological correspondence between the EELRs and the radio structure. This fact indicates a mechanism connected with the production of the radio jet that is much less strongly collimated than the jet itself. We have suggested in this context that the initiation of an FR\,II jet is accompanied by a nearly hemispherical blast wave \citep{Fu07b}. In the usual case of two jets, these blast waves can be capable, at least in some cases, of clearing most of the interstellar medium from the host galaxy. Although, as we have emphasized, 3C\,48 is atypical in several ways, it is not unreasonable to suppose that it gives us a fairly typical picture of a very young radio jet; and in this case we see a massive, high-velocity, wide-solid-angle outflow coming from a region near the base of the radio jet, tending to corroborate the picture we have suggested for the FR\,II quasars with luminous EELRs.

\acknowledgments 
H. F. is grateful for the hospitality of the Purple Mountain Observatory, where part of this work was completed. We thank Luis Ho, David Rupke, Jong-Hak Woo, Yanling Wu, and Lin Yan for helpful discussions. We also thank the anonymous referee for helping us to improve the paper and clarify some obscure points. This work is based on observations made with the Spitzer Space Telescope, which is operated by the Jet Propulsion Laboratory, California Institute of Technology, under a contract with NASA. Support for this work was provided by NASA through an award issued by JPL/Caltech and by NSF under grant AST-0807900. This research has made use of the NASA/IPAC Extragalactic Database (NED) which is operated by the Jet Propulsion Laboratory, California Institute of Technology, under contract with the National Aeronautics and Space Administration.

%%%%%%%%%%%%
% Figures
%%%%%%%%%%%%
\clearpage

\end{document}